# Performance Analysis of Non-Orthogonal Multiple Access in Free Space Optical Communication System

Ruijie Li and Anhong Dang

*Abstract*—In this paper, a multipoint-to-point system consisting $K$ users and a central node with non-orthogonal multiple access (NOMA) technique over free space optical (FSO) channel is characterized for two cases. In the case of guaranteed user's quality of service (QoS), the outage probability is derived. Another case is that user's QoS is determined by their own channel conditions. The developed analytical results of ergodic sum data rate show that NOMA outperforms orthogonal multiple access (OMA). Besides, the power control scheme is proposed, and results show that the power control scheme has great influence to the user's QoS. Monto Carlo simulation has been done, and it matches quite well with the theoretical analysis.

## I. INTRODUCTION

Driven by the tablets, smartphones, and real-time bandwidth-intensive applications, wireless traffic will increase by over a factor of 100: form under 3 exabytes in 2010 to over 190 exabytes by 2018 [1]. Free space optical (FSO) communication has attracted much interest in both academia and industry as a high-speed wireless communication technology, due to the advantage of quick and easy deployment, high bandwidth qualities, and high security [2, 3]. Especially when the radio-frequency (RF) spectrum is heavily congested, the features of no need for license fees and no government regulations restricting the use of bandwidth become the significant advantages [4-6].

As a wireless technology, it is imperative that FSO communication can support multiple users to access the network. Generally, the multiple access techniques include time division multiple access (TDMA), frequency division multiple access (FDMA), and code division multiple access (CDMA). In [7], authors proposed an optical CDMA (OCDMA) network by assigning the fast frequency hopping-based codes. However, for OCDMA techniques, the optical orthogonal codes (OOCs) are considered, which must use long, sparse codes to achieve acceptable performance, hence, the resource reuse is limited [8]. The TDMA is considered in [9], which means only one user can be served in each time instant. The space and time division multiple access technique is proposed by J. Liu etc., in which the space dimension is utilized [10]. However, most of these works refer to the orthogonal multiple access (OMA), which cannot provide sufficient resource reuse.

For further enhance system capacity and provide enhanced user's quality of service (QoS), non-orthogonal multiple access (NOMA) has recently been proposed [11-14], in which users are allocated in power domain to realize multiple access. Therefore, signals from different users are allocated in the same time-frequency domain. After receiving the data at the receiver, successive interference cancellation (SIC) is carried out at the receiver side for signal detection, which means channel state information (CSI) is required. For FSO communication system, the turbulence channel is a typical slow fading channel (the correlation time is on the order of millisecond [4]), CSI can be estimated easily and accurately. To the best of our knowledge, a performance analysis of NOMA in FSO communications lacks the open literature so far.

In this paper, we consider a multipoint-to-point FSO system, consisting of several users and a central node, shown in Fig. 1. Inspired by the basic idea of NOMA, each user can communicate with the central node by using same time-frequency resource. Because the channel fading of different users are independent, each user's message experiences distinct channel fading. Atmospheric channel fading is no longer treated as the nuisance which has to be overcome, but as a source of randomization that the central node can split them in power domain though SIC. Meanwhile, a power back-off scheme is proposed to make sure that different users can be distinguished effectively. The performance of the system is evaluated in two types of situations just like in [12, 13]. Firstly, we consider the case where each user has a targeted data rate. The outage probability is an ideal metric for performance evaluation because it measures the ability that the system can maintain the user's QoS. Theoretical and simulation results provide that the outage probability performance is influenced by not only the targeted data rate but also the transmitted power of each user. Secondly, we consider the case where use's QoS is determined by their own channel conditions. The ergodic sum rate is evaluated, and we show that NOMA outperforms OMA. And

The authors are with the School of Electronics Engineering and Computer Science, Peking University, Beijing 100871, China (email: liruijie@pku.edu.cn; ahdang@pku.edu.cn).

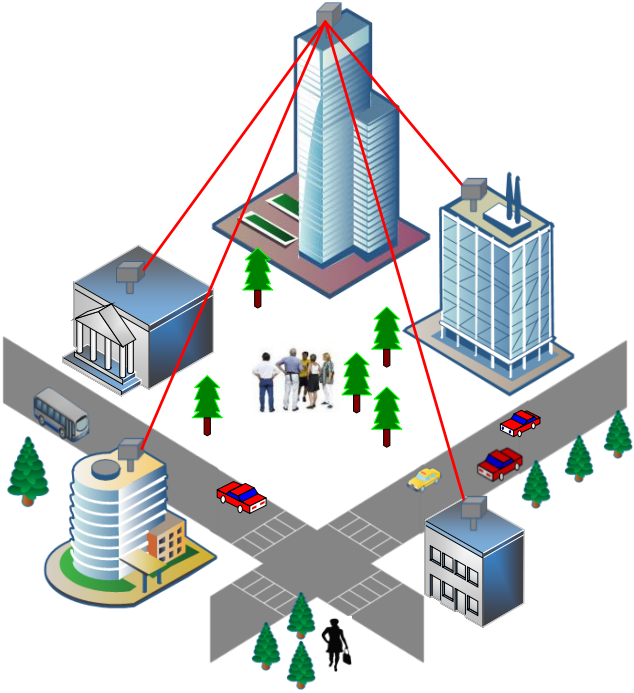

Fig. 1. Multipoint-to-point FSO communication system.

also, the performance loss of NOMA is in significant with the increase of turbulence strength.

This paper is organized as follows. The system model of the proposed FSO network is introduced in Section II. In Section III, the theoretical analyses for the two types of situation are shown. The simulation results are depicted in Section IV. Finally, conclusions are drawn.

## II. SYSTEM MODEL

In this paper, we consider a multipoint-to-point FSO communication system consisting of several transmitters (i.e. different users) and an optical receiver (i.e. central node), over FSO channel (shown in Fig. 1). We assume that each user is equipped with one aperture for transmit the optical signal, and at the receiver end only one aperture is used for receiving the signal. For this model, the users transmit their own signal on the same time-frequency resources. The central node decodes the signal by utilizing SIC on the power domain. In the following, we assume that the users to be served quasi-static, so that their CSI is not outdated until the next channel estimation. Without loss of generality, assuming that all the users are ordered based on their channel gain

$$g_1 \geq \cdots \geq g_k \cdots \geq g_K, \quad (1)$$

where $g_k$ is the channel fading of the $k$-th user, $K$ is the number of users in the system. In this system, the channel gain is considered to be a product of two factors, i.e. $g_k = L_k I_k$, where $L_k$ is the deterministic propagation loss and $I_k$ is the intensity scintillation caused by turbulence. $L_k$ is determined by the exponential Beers-Lambert law as [15, 16]

$$L_k = e^{-\Phi \times d_k}, \quad (2)$$

where $d_k$ is the link distance between $k$-th user and central node, and $\Phi$ is the atmospheric attenuation coefficient, which is given by [15]

$$\Phi = \frac{3.91}{V(\text{km})} \times \left(\frac{\lambda(\text{nm})}{550}\right)^{-q}, \quad (3)$$

where $V$ is the visibility in kilometers, $\lambda$ is the wavelength of laser in nanometers, and $q$ is a parameter related to the visibility, being $q = 1.3$ for average visibility (6 $km$<V<50 $km$). $q = 0.16 V + 0.34$ for haze visibility (1 $km$<V<6 $km$). For the intensity scintillation, gamma-gamma probability density function (PDF) is the most widely accepted model for describing the distribution of intensity scintillation, which is [17]

$$f_I(I) = \frac{2(\alpha\beta)^{(\alpha+\beta)/2}}{\Gamma(\alpha)\Gamma(\beta)} I^{(\alpha+\beta)/2-1} K_{\alpha-\beta}\left(2\sqrt{\alpha\beta I}\right), \quad I > 0, \quad (4)$$

where $\Gamma(\cdot)$ is the gamma function, $K_{\alpha-\beta}(\cdot)$ is the ($\alpha$-$\beta$)th-order modified Bessel function, $\alpha$ and $\beta$ are determined by Rytov variance ($\sigma_R^2$) related with the atmospheric conditions [17]

$$\alpha = \left[\exp\left(\frac{0.49\sigma_R^2}{\left(1+1.11\sigma_R^{12/5}\right)^{7/6}}\right) - 1\right]^{-1}, \quad (5)$$

$$\beta = \left[\exp\left(\frac{0.51\sigma_R^2}{\left(1+0.69\sigma_R^{12/5}\right)^{5/6}}\right) - 1\right]^{-1}. \quad (6)$$

Intensity modulation/direct detection (IM/DD) scheme is considered. Assuming that the transmitted signal for $k$-th user is $x_k = P_k s_k$, where $s_k$ is the transmitted data, and $P_k$ is the transmitted power for $k$-th user. Then, the received signal at the central node is

$$y = \sum_{i=1}^{K} g_i x_i + n = \sum_{i=1}^{K} g_i P_i s_i + n, \quad (7)$$

where $n$ is the additive noise, modeled as additive white Gaussian noise (AWGN), with zero mean and variance of $\sigma^2$. A constant noise power spectral density (PSD), denoted by $N_0$, is assumed so that $\sigma^2 = N_0 B$, where $B$ is the signal bandwidth. At the receiver, SIC is adopted. When detects $k$-th user's messages, it already decodes the prior $i$-th ($i<k$) user's message, then remove the message from its observation, in a successive manner. The rest ($K$-$k$)-th messages are regarded as interferences. Based on the [Eq. (46), [18]], the achievable data rate for $k$-th user is given below when SIC is considered.

$$R_k = \begin{cases} \left[\frac{1}{2}\log\left(1+\frac{(\mu_k g_k P_k)^2}{\sum_{i=k+1}^{K}(\mu_i g_i P_i)^2 + A\sigma^2}\right) - \varepsilon_\phi\right]^+ \geq \tilde{R}_k, & k < K \\ \left[\frac{1}{2}\log\left(1+\frac{(\mu_K g_K P_K)^2}{A\sigma^2}\right) - \varepsilon_\phi\right]^+ \geq \tilde{R}_K, & k = K \end{cases} \quad (8)$$

where $\tilde{R}_k$ is the targeted data rate of $k$-th user, with $A=9(1+\varepsilon_\mu)^2$, $\varepsilon_\varphi=0.016$ and $\varepsilon_\mu=0.0015$. $[x]^+$ denotes max$\{0,x\}$, and $\mu_i \in [0,0.5]$ represents the ratio between expectation of the received power and the maximum received power [18]. If (8) is satisfied, we assume that perfect SIC can be performed in the central node without error propagations.

## III. PERFORMANCE ANALYSIS

### A. Transmission Power Control

In point-to-point FSO system the transmitted power $P$ can be expressed as

$$P = P_{aim}/L, \quad (9)$$

where $P_{aim}$ is the target arrived power and $L$ is the power loss caused by path loss. At the central node, SIC technique requires diverse arrived power to distinguish multiplexing users. To obtain it, besides the randomization channel fading caused by turbulence channel, a power back-off scheme is considered. Inspired by [14], the transmitted power of $k$-th user in the system is expressed as

$$P_k = \frac{P_{aim}}{L_k 10^{\frac{(k-1)\varsigma}{10}}} = a_k P_{aim}, \quad (10)$$

where $\varsigma$ is the power back-off step in dB. This scheme indicates that the arrived power of users is gradually degraded with a step of $\varsigma$ dB. Although there is no turbulence during the link, the second strongest use's power is $\varsigma$ dB lower than that of the strongest user, and so on. Assuming that $i$-th user is prior to $k$-th user

$$g_i P_i > g_k P_k. \quad (11)$$

Substituting (10) into (11)

$$I_i > I_k 10^{\frac{(i-k)\varsigma}{10}}. \quad (12)$$

Since $\varsigma$ is bigger than 0 dB, we consider $I_i > I_k$ as a condition to cover all possible values of $\varsigma$.

### B. Distribution Function of the FSO Channel

Denotes $h_k = I_k^2$. The PDF of the unordered channel fading can be calculated using the "change of variable" method, therefore, the PDF of $h_k$ can be expressed as

$$f_{h_k}(h) = \frac{1}{\sqrt{h}} \frac{(\alpha\beta)^{(\alpha+\beta)/2}}{\Gamma(\alpha)\Gamma(\beta)} \left(\sqrt{h}\right)^{(\alpha+\beta)/2-1} K_{\alpha-\beta}\left(2\sqrt{\alpha\beta\sqrt{h}}\right). \quad (13)$$

Integral (13), the cumulative distribution function (CDF) of the unordered variable $h_k$ can be expressed as

$$F_{h_k}(y) = \frac{(\alpha\beta)^{\frac{\alpha+\beta}{2}} y^{\frac{\alpha+\beta}{4}}}{4\pi\Gamma(\alpha)\Gamma(\beta)} G_{1,5}^{4,1}\left(\frac{1}{16}(\alpha\beta)^2 y \left| \begin{array}{c} 1-(\alpha+\beta)/4 \\ \frac{\alpha-\beta}{4}, \frac{\alpha-\beta+2}{4}, \frac{\beta-\alpha}{4}, \frac{2-\alpha+\beta}{4}, -\frac{\alpha+\beta}{4} \end{array}\right.\right), \quad (14)$$

where $G_{p,q}^{m,n}(y|\cdot)$ is the Meiger's G-function [19]. Deriving from (12), $h_1 \geq \cdots \geq h_k \cdots \geq h_K$. Using order statistics [20], the PDF of the ordered variable $h_k$ is

$$f'_{h_k}(h) = \frac{K!}{(k-1)!(K-k)!}\{F_{h_k}(h)\}^{K-k}\{1-F_{h_k}(h)\}^{k-1} f_{h_k}(h). \quad (15)$$

The joint density of all $k$-order statistics is [20]

$$f'_{h_1,\cdots,h_k}(h_1,\ldots,h_k) = k! \prod_{i=1}^{k} f_{h_i}(h_i), h_1 \geq \cdots \geq h_k. \quad (16)$$

### C. Case 1: Outage Probability for Maintained QoS

Considering each user has a targeted data rate, which is determined by their own QoS. When constrains are satisfied, i.e. $R_k \geq \tilde{R}_k$, the sum rate of the network is $\sum_k \tilde{R}_k$. Therefore, the sum rate is not of interest in this case. The outage probability is an ideal metric for performance evaluation. Base on (8), the outage probability can be expressed as

$$P_k^{out} = 1 - \mathbb{P}(E_1^c \cap \cdots \cap E_k^c) = 1 - \prod_{i=1}^{k} \mathbb{P}(E_i^c), \quad (17)$$

where $E_k \triangleq \{R_k < \tilde{R}_k\}$, and $E_k^c$ is the complementary set of $E_k$. When $k<K$,

$$\mathbb{P}(E_k^c) = \mathbb{P}\left\{\left[\frac{1}{2}\log\left(1 + \frac{(\mu_k g_k P_k)^2}{\sum_{i=k+1}^{K}(\mu_i g_i P_i)^2 + A\sigma^2}\right) - \varepsilon_\phi\right]^+ \geq \tilde{R}_k\right\}$$

$$\stackrel{(a)}{=} 1 - \mathbb{P}\left\{h_k \leq \frac{A\phi_k + \phi_k \rho \sum_{i=k+1}^{K} L_i^2 \mu_i^2 h_i a_i^2}{\rho \mu_k^2 a_k^2 L_k^2}\right\}$$

$$= 1 - \int \cdots \int_{y_{k+1}}^{\upsilon} f'_{h_k,\cdots,h_K}(y_k,\ldots,y_K) dy_k \cdots dy_K, \quad (18)$$

where $\upsilon = \left(A\phi_k + \phi_k \rho \sum_{i=k+1}^{K} L_i^2 \mu_i^2 h_i a_i^2\right) \Big/ \left(\rho \mu_k^2 a_k^2 L_k^2\right)$, $\phi_k = e^{2(\tilde{R}_k + \varepsilon_\phi)} - 1$, and $\rho = P_{aim}^2/N_0 B$. Step (a) is obtained by assuming $h_k/\upsilon > \exp(2\varepsilon_\phi) - 1$, which can be satisfied for most of the situations. If $\upsilon \geq h_{k+1}$, substituting (16) into (18)

$$\mathbb{P}(E_k^c) = 1 - \int \cdots \int_{y_{k+1}}^{\upsilon} f'_{h_k,\cdots,h_K}(y_k,\ldots,y_K) dy_k \cdots dy_K$$

$$= 1 - \int \cdots \int_{y_{k+1}}^{\upsilon} \left[\int_{y_k}^{+\infty} \cdots \int_{y_2}^{+\infty} f'_{h_k,\cdots,h_K}(y_1,\ldots,y_K) dy_1 \cdots dy_{k-1}\right] dy_k \cdots dy_K$$

$$= 1 - \int \cdots \int_{y_{k+1}}^{\upsilon} \frac{K!}{(k-1)!} \prod_{i=k}^{K} f_{h_i}(y_i)[1-F(y_k)]^{k-1} dy_k \cdots dy_K$$

$$= \frac{K!}{k!} \int_0^{+\infty} \cdots \int_{y_{k+2}}^{+\infty} \prod_{i=k+1}^{K} f_{h_i}(y_i)[1-F(\upsilon)]^k dy_{k+1} \cdots dy_K. \quad (19)$$

When $k=K$, (18) is changed to

$$\mathbb{P}\left(E_K^c\right) = 1 - \mathbb{P}\left\{\frac{\rho(\mu_K g_K a_K)^2}{A} \leq \phi_K\right\}$$
$$= 1 - \int_0^{\psi_K} K\left[1 - F_{h_k}(y)\right]^{K-1} f_{h_k}(y) dy \quad (20)$$
$$= \left[1 - F_{h_k}(\psi_K)\right]^K,$$

where $\psi_K = A\phi_K / \rho\mu_K^2 L_K^2 a_K^2$. Substituting (18) and (20) into (17), the outage probability of $k$-th user can be derived.

The coverage probability is defined as the probability that all the users in the system can achieve reliable detection, given by

$$P^{cov} = \prod_{k=1}^{K}\left(1 - P_k^{out}\right). \quad (21)$$

### D. Case 2: Ergodic Sum Rate

In this case, user's QoS is determined by their own channel conditions, i.e. $R_k = \tilde{R}_k$. In this scenario, (8) always holds, and all users can be served with zero outage probability but with different data rate. The ergodic sum rate is evaluated, and can be expressed as

$$R_{sum}^{NOMA} = E\left\{\sum_{k=1}^{K} R_k\right\}$$
$$\stackrel{(b)}{=} E\left\{\frac{1}{2}\log\left(1 + \frac{\rho\sum_{k=1}^{K}(\mu_k g_k a_k)^2}{A}\right) - K\varepsilon_\phi\right\}, \quad (22)$$

where $E\{\cdot\}$ is the expectation over channel fading. Step (b) is obtained by substituting (8) into it.

## IV. SIMULATION RESULTS

In this section, the performance of NOMA is evaluated, where OMA is used for benchmarking. Both theoretical and Monto Carlo simulation results are presented, denoted as "Theo." and "Simu.", respectively. Two user system is considered and the link distances between users and central node are 1 *km* and 3 *km*, respectively. The wavelength of the laser is 1550 nm. Clear visibility of 16 *km* with $\sigma_R^2 = \{0.1, 1\}$ is considered. In Fig. 2, the outage probability for two uses are depicted when the power back-off are 2 dB, 3 dB, and 5 dB. The Rytov variance ($\sigma_R^2$) is 0.1, and the targeted data rates for two users are $\tilde{R}_1=0.5$ and $\tilde{R}_2=0.5$, respectively. The outage performance of 5 dB outperforms that of 2 dB and 3 dB power back-off because large power back-off helps to decrease the interference from the second user. However, user 2 holds the opposite results. The reason is that even better performance of the first user helps to decode user 2, the arrived signal-to-noise ratio (SNR) of second user is too small when the back-off power is big enough. Figure 3 presents the similar results as Fig. 2, when the Rytov variance is 1. Comparing Fig. 2 with Fig. 3, the strength of turbulence has great influence to the outage probability performance as expected. In both figures, the Monto Carlo results match quite well with the theoretical results. In Fig. 4, the outage probability performances of two users under different targeted data rate when $\sigma_R^2=0.1$ are shown. Notice that higher targeted data rate results in worse outage probability as it is more difficult to be satisfied. Fig. 5 presents the outage performance under $\sigma_R^2=1$. It shows the similar results as in Fig.4. Furthermore, the theoretical results match the simulation quite well. In Fig. 6, the ergodic sum rate between NOMA and OMA are presented as the function of $\rho$ when $\zeta=5$dB. Figure 6 demonstrated that the NOMA can achieve a larger sum rate than OMA under different turbulence strength. The performance loss of NOMA is quite small compared with OMA with the increase of turbulence strength. Besides, the advantage of NOMA at low SNR is insignificant.

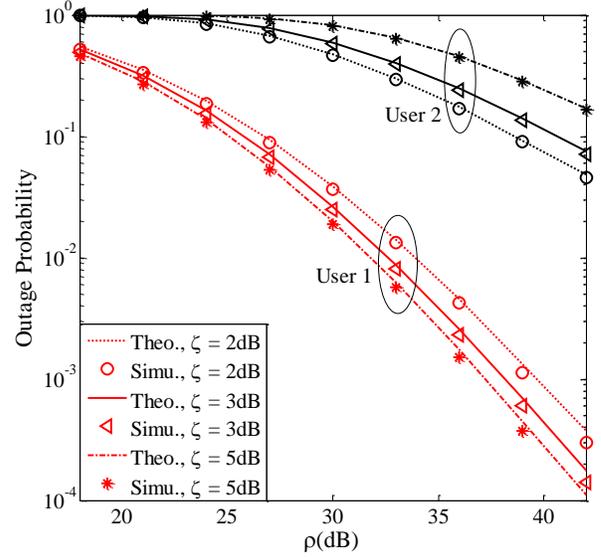

Fig. 3. Outage probability of NOMA for different $\zeta$ when $\sigma_R^2=1$.

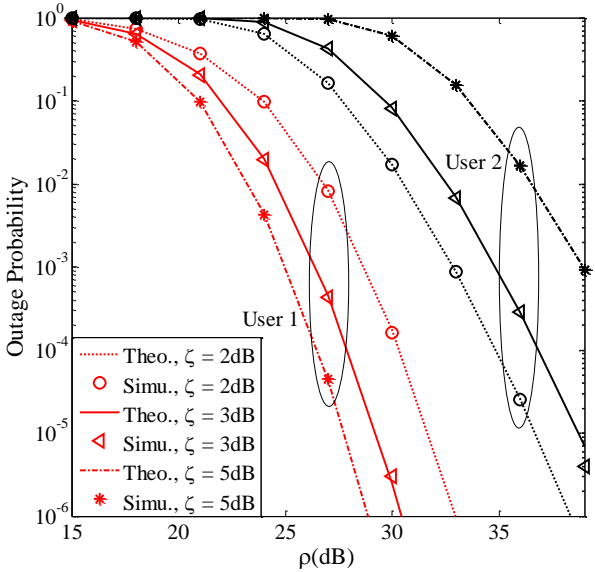

Fig. 2. Outage probability of NOMA for different $\zeta$ when $\sigma_R^2=0.1$.

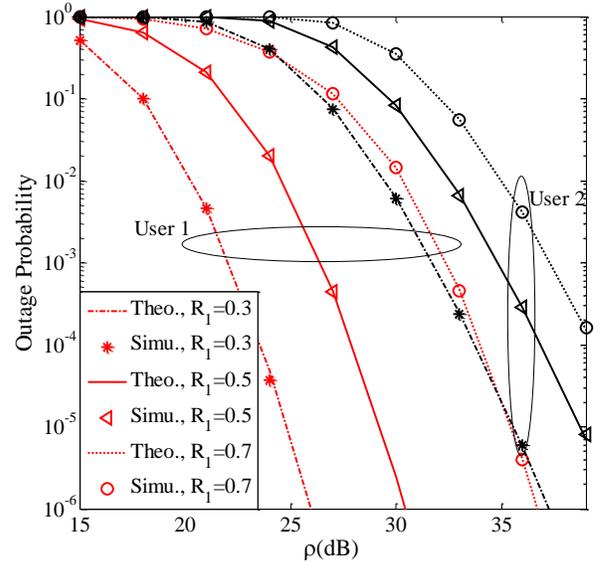

Fig. 4. Outage probability of NOMA for different targeted data rate when $\sigma_R^2$ is 0.1.

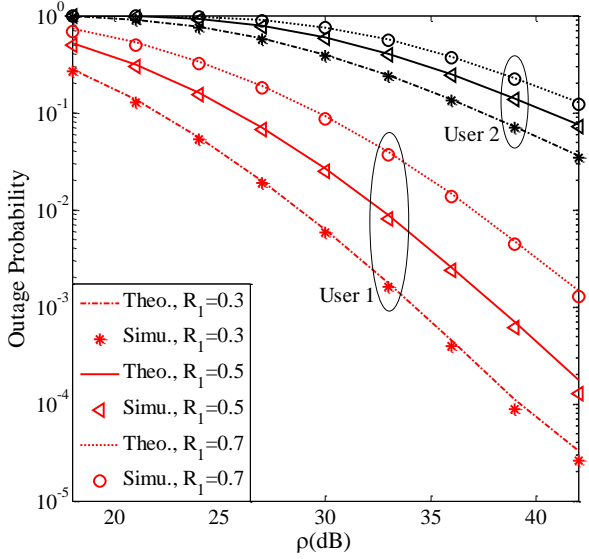

Fig. 5. Outage probability of NOMA for different targeted data rate when $\sigma_R^2$ is 1.

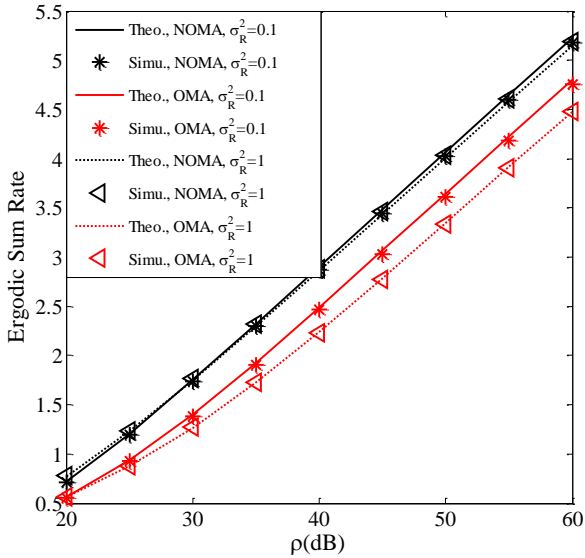

Fig. 6. Ergodic sum data rate performances of NOMA and OMA.

## V. CONCLUSIONS

In this paper, NOMA techniques with power control scheme has been proposed for FSO communication system. The performance in terms of outage probability and ergodic sum data rate were theoretically analyzed. We have shown that the back-off power control has great influence to the user's QoS. NOMA achieves a superior ergodic sum data rate than OMA. NOMA is more suitable for FSO link, especially for strong turbulence. However, the performance gain is insignificant for low SNR. Because SIC has to be considered, the proposed system introduces additional complexity. It is quite important to achieve a tradeoff between performance and complexity. Monto Carlo simulations match quite well with the theoretical results.